\documentclass[twocolumn]{aastex63}
\usepackage{graphics,epsf}
\usepackage{amsmath}                
\usepackage{amsfonts}               
\usepackage{amssymb}                
\usepackage{epsfig}                 
\usepackage{appendix}
\usepackage{graphicx}
\usepackage{float}
\usepackage{color}
\usepackage{multirow}
\usepackage{colortbl}
\usepackage[para,online,flushleft]{threeparttable}

\hypersetup{citecolor=blue, 
            linkcolor=red, 
            menucolor=blue, 
            urlcolor=blue}  

\newcommand{\cm}{{~\rm cm}}
\newcommand{\m}{{~\rm m}}

\newcommand{\s}{{~\rm s}}
\newcommand{\h}{{~\rm h}}
\newcommand{\msec}{{~\rm ms}}
\newcommand{\g}{{~\rm g}}
\newcommand{\G}{{~\rm G}}
\newcommand{\K}{{~\rm K}}
\newcommand{\erg}{{~\rm erg}}

\begin{document}

\title{Powering luminous core collapse supernovae with jets}


\author[0000-0003-0375-8987]{Noam Soker}
\affiliation{Department of Physics, Technion, Haifa, 3200003, Israel;  soker@physics.technion.ac.il}

\begin{abstract}
I examine recent fittings of luminous supernovae (LSNe) with extra energy sources of magnetar and helium burning and find that in about half of these LSNe the fitting parameters have some problems. In some LSNe the total energy of these two energy sources is larger than the kinetic energy of the ejecta that the fitting yields. In some others LSNe the total energy of the delayed neutrino explosion mechanism and these two extra sources combined is smaller than the kinetic energy that the fitting yields. These difficulties suggest that, like earlier claims that jets power superluminous supernovae (SLSNe), jets also power the less luminous LSNe. A magnetar might also supply energy. However, in most cases jets supply more energy than the magnetar, during the explosion and possibly at late times. I strengthen an earlier claim that jets launched at magnetar birth cannot be ignored. I explain the trend of maximum rise time for a given luminosity of hydrogen deficient core collapse supernovae (CCSNe), in particular LSNe and SLSNe, with a toy model of jets that are active for a long time after explosion. 
 \end{abstract}


\section{Introduction} 
\label{sec:intro}

Recent studies concentrate on two core collapse supernova (CCSN) explosion mechanisms that utilise the gravitational energy that the core of a massive star releases as it collapses to form a neutron star (NS). 
These are the delayed neutrino mechanism (\citealt{BetheWilson1985}; for later studies see, e.g.,  \citealt{Hegeretal2003, Janka2012, Nordhausetal2012, CouchOtt2013, Bruennetal2016, Jankaetal2016R, OConnorCouch2018, Mulleretal2019Jittering, BurrowsVartanyan2021, Fujibayashietal2021, Bocciolietal2022, Nakamuraetal2022}) and the jittering jets explosion mechanism (\citealt{Soker2010}; for later papers see, e.g.,  \citealt{PapishSoker2014Planar, GilkisSoker2015, Quataertetal2019, Soker2019RAA, Soker2020RAA, AntoniQuataert2022, ShishkinSoker2022, Soker2022a, Soker2022Boosting}).   
 
The basic properties of the jittering jets explosion mechanism distinguish it from jet-driven explosion mechanisms that require the pre-collapse core to have fast rotation (e.g., \citealt{Khokhlovetal1999, Aloyetal2000, MacFadyenetal2001, Maedaetal2012, LopezCamaraetal2013, BrombergTchekhovskoy2016,  Nishimuraetal2017, Sobacchietal2017, Grimmettetal2021, Gottliebetal2022, Perleyetal2022}). These properties are as follows.  (1) The jittering jets explosion mechanism is based on a negative feedback mechanism (see review by \citealt{Soker2016Rev}). This holds also for cases where the pre-collapse core is rapidly rotating and therefore jittering angles are very small, i.e., the jets-axis is more or less constant. (2) The jittering jets explosion mechanism attributes the powering of most (or possibly all) CCSNe to jets, including cases of non-rotating pre-collapse cores. If the pre-collapse core rotates very slowly then the stochastic convective motion in the core (e.g., \citealt{GilkisSoker2014, GilkisSoker2016, ShishkinSoker2021}) or in the envelope (e.g., \citealt{Quataertetal2019}) allows the formation of intermittent/stochastic accretion belts or accretion disks. The accretion disks/belts launch the jittering jets. As a result of that according to the jittering jets explosion mechanism there are no failed CCSNe (e.g., \citealt{Gilkisetal2016Super, Soker2017RAA, AntoniQuataert2022}), a claim that the observational finding by \cite{ByrneFraser2022} supports. Many CCSN remnants that have morphological features that are imprints of jets (e.g., \citealt{Bearetal2017, GrichenerSoker2017, YuFang2018, Luetal2021, Soker2022a}) also support the notion that jets power most CCSNe.

When the jets maintain a more or less fix axis, namely, for cases with rapid pre-collapse core rotation, the jets explode the core and envelope mass along and near the polar directions very efficiently. However, the efficiency of ejecting the stellar zones in and near the equatorial plane is inefficient. The accretion of equatorial gas feeds the newly born NS that might grow to a black hole. The accretion is via an accretion disk around the NS and later possibly around the black hole. This accretion flow launches jets that power an energetic CCSN. 
This implies that according to the jittering jets explosion mechanism, or more generally the jet feedback explosion mechanism, the formation of a black hole comes along with an energetic CCSN rather than a failed CCSN(e.g., \citealt{Gilkisetal2016Super, Soker2017RAA}). 

Consider superluminous supernovae (SLSNe), i.e., those with peak r-band magnitude of $M_{\rm r} < -20$, and luminous supernovae (LSNe) that have $M_{\rm r} =-19$ to $-20$ (see definition by \citealt{Gomezetal2022}). 
Because the Neutrino driven explosion mechanism cannot explain CCSNe with explosion energies of $E_{\rm SN} \ga 2 \times 10^{51} \erg$ (e.g., \citealt{Fryer2006, Fryeretal2012, Papishetal2015a, Sukhboldetal2016, SukhboldWoosley2016, Gogilashvilietal2021}), models of SLSNe that apply the delayed neutrino explosion mechanism require extra energy sources, such as a magnetar (e.g., \citealt{Greineretal2015, Metzgeretal2015, Yuetal2017, Marguttietal2018}). 
Other studies \citep{Soker2016Magnetar, Soker2017Magnetar, SokerGilkis2017} conclude that energetic magnetars are accompanied by energetic jets in the explosion, and possibly also by late jets, and that in many cases the jets carry more energy than the magnetar does (for a review of possible energy sources of SLSNe see, e.g., \citealt{WangWangDai2019RAA}.) 
In particular, \cite{SokerGilkis2017} conclude that for about half of the 38 SLSNe that \cite{Nicholletal2017b} model with a magnetar the explosion energy itself has an energy of 
$E_{\rm SN}  > 2 \times 10^{51} \erg$. This is more than what the delayed neutrino mechanism can supply. Their conclusion, which strengthens the earlier two papers cited above \citep{Soker2016Magnetar, Soker2017Magnetar}, is that jets must play a crucial role in the powering of SLSNe. In a recent paper \cite{Reichertetal2022} further support this claim by conducting magnetohydrodynamic simulations. 

In the present paper I examine the recent study by \cite{Gomezetal2022} who model 40 LSNe with extra powering by a magnetar (section \ref{sec:Magnetar}). Motivated by my finding that jets must be included in more than half of these LSNe (section \ref{sec:Problems}), and possibly in all of them, I examine one particular property of SLSNe when I consider the powering to be by late jets. I end with my conclusion that LSNe, like SLSNe, are also powered by jets (section \ref{sec:Summary}).  

\section{The magnetar and nickel powering MOSFiT model} 
\label{sec:Magnetar}
   
\cite{Gomezetal2022} fit the light curve of 40 LSNe by considering extra powering (in addition to the explosion itself) by a magnetar and by nickel decay. They use the MOSFiT package to determine the best parameters that fit the light curve. 
\cite{Gomezetal2022} present the values of the different variables of their fitting, including the uncertainties. I present the best values of some of these quantities (without the uncertainties) in Table \ref{Table:fit}. In the first column I list the names of the 40 LSNe, and from the second to fifth columns I present their best values for the kinetic energy of the ejecta $E_{\rm kin}$, the mass of the ejecta $M_{\rm ej}$, the $^{56}{\rm Ni}$ mass $M_{\rm Ni}$, and the initial spin period of the magnetar $E_{\rm spin}$. For the NS mass they consider the range of $M_{\rm NS}=1.5 M_\odot$ to $M_{\rm NS}=1.9 M_\odot$. 
\begin{table*}
\centering
\begin{tabular}{|c|c|c|c|c|c|c|c|c|}
\hline
Name & $E_{\rm kin}$ & $M_{\rm ej}$ & $M_{\rm Ni}$ & $P_{\rm spin}$ & $E_{\rm mag}(1.7)$ & $\eta_{\rm min}$ & $\eta_{\rm max}$ & Problems \\ \hline
 & ($10^{51} \erg$) & ($M_\odot$ ) & ($M_\odot$ ) & (ms) & ($10^{51} \erg$) & (eq. \ref{eq:EtaMin}) & (eq. \ref{eq:EtaMax}) & \\ \hline
DES14Clrhg & 1.3 & 4.1 & 0.04 & 10 & 0.35 & 0.22 & {1.96} &  \\ \hline
DES15C3hav & 1.7 & 8.1 & 0.08 & 2.4 & 6.04 & 2.9 & {5.53} & Excess-E \\ \hline
DES16C3cv  & 1.3 & 5.2 & 0.06 & 4.2 & 1.97 & 1.3 & {3.48} & Excess-E \\ \hline
iPTF13dnt  & 11.8 & 9.5 & 1.28 & 20 & 0.09 & 0.006 & {0.54} & Missing-E  \\ \hline
iP TF16asu & 0.04 & 0.4 & 0.003 & 13.2 & 0.2 & 4.1 & {56.15} & Excess-E \\ \hline
iP TF17cw  & 1.8 & 4.3 & 0.14 & 22.5 & 0.07 & 0.03 & {1.41} &  \\ \hline
OGLE15xl   & 8.7 & 10.5 & 1.65 & 10.9 & 0.29 & 0.03 & {0.9} & Missing-E  \\ \hline
PS15cvn    & 0.8 & 3.7 & 0.14 & 19.1 & 0.1 & 0.10 & {3.22} &  \\ \hline
PTFl0gvb   & 5 & 4.2 & 0.31 & 23.2 & 0.06 & 0.01 & {0.62} & Missing-E \\ \hline
PTF10iam   & 0.05 & 0.9 & 0.005 & 13.7 & 0.19 & 3.1 & {44.71} & Excess-E \\ \hline
PTF11img   & 6.5 & 3.9 & 0.84 & 18.5 & 0.1 & 0.01 & {0.75} & Missing-E \\ \hline
PTF12gty   & 1.9 & 9.6 & 0.18 & 7.4 & 0.64 & 0.28 & {1.76} &  \\ \hline
PTF12hni   & 4.8 & 14.1 & 0.35 & 5 & 1.39 & 0.24 & {1} &  \\ \hline
SN1991D    & 3.6 & 5.3 & 0.2 & 3.1 & 3.62 & 0.83 & {1.93} &  \\ \hline
SN2003L    & 10.4 & 17.2 & 1.97 & 16.6 & 0.13 & 0.01 & {0.83} & Missing-E \\ \hline
SN2007ce   & 2 & 10.1 & 0.21 & 20.9 & 0.08 & 0.03 & {1.39} &  \\ \hline
SN2009cb   & 4.3 & 7.1 & 0.21 & 2.8 & 4.44 & 0.86 & {1.85} &  \\ \hline
SN2010ay   & 3.7 & 8.1 & 1.19 & 20.7 & 0.08 & 0.02 & {1.63} &  \\ \hline
SN2011kl   & 12.1 & 3.9 & 1.06 & 11 & 0.29 & 0.02 & {0.48} & Missing-E \\ \hline
SN2012aa   & 1.1 & 26.7 & 0.09 & 6.2 & 0.91 & 0.68 & {3.06} &  \\ \hline
SN 2013hy  & 3.8 & 17 & 1.76 & 14 & 0.18 & 0.04 & {2.11} &  \\ \hline
SN2018beh  & 2.1 & 10.6 & 0.03 & 1.5 & 15.46 & 6.1 & {9.7} & Excess-E \\ \hline
SN2018don  & 1.2 & 6.7 & 0.02 & 7.8 & 0.57 & 0.40 & {2.28} &  \\ \hline
SN2018fcg  & 1 & 2.1 & 0.01 & 6.3 & 0.88 & 0.73 & {3.07} &  \\ \hline
SN2019cri  & 2.1 & 25.6 & 0.52 & 12.9 & 0.21 & 0.08 & {1.89} &  \\ \hline
SN 2019dwa & 1.2 & 2.7 & 0.04 & 12.5 & 0.22 & 0.15 & {2} &  \\ \hline
SN2019gam  & 36.3 & 29.5 & 2.15 & 3.1 & 3.62 & 0.08 & {0.37} & Missing-E  \\ \hline
SN2019hge  & 1.3 & 16.7 & 0.1 & 1.7 & 12.04 & 7.7 & {12.73} & Excess-E \\ \hline
SN2019J    & 1.3 & 13.7 & 0.03 & 1.2 & 24.16 & 15.4 & {23.57} & Excess-E \\ \hline
SN 2019moc & 1.7 & 4.4 & 0.43 & 19.9 & 0.09 & 0.04 & {2.07} &  \\ \hline
SN2019obk  & 4.2 & 14.5 & 0.55 & 8.8 & 0.45 & 0.09 & {1.03} &  \\ \hline
SN2019pvs  & 1.9 & 16.2 & 0.58 & 9.1 & 0.42 & 0.18 & {2.32} &  \\ \hline
SN 2019stc & 1 & 4 & 0.08 & 7.7 & 0.59 & 0.49 & {2.96} &  \\ \hline
SN2019unb  & 1.5 & 15.2 & 0.17 & 1.9 & 9.64 & 5.3 & {9.3} & Excess-E \\ \hline
SN2019uq   & 2.3 & 7.9 & 0.23 & 17 & 0.12 & 0.04 & {1.26} &  \\ \hline
SN2019wpb  & 4.1 & 7.4 & 0.63 & 21.5 & 0.08 & 0.02 & {1.02} &  \\ \hline
SN 2021lei & 8.6 & 7.8 & 0.74 & 20.1 & 0.09 & 0.008 & {0.53} & Missing-E \\ \hline
SN2021lwz  & 0.02 & 0.4 & 0.002 & 11 & 0.29 & 11.9 & {117.32} & Excess-E \\ \hline
SN2021uvy  & 0.7 & 3.8 & 0.02 & 5.9 & 1 & 1.2 & {4.64} & Excess-E \\ \hline
SN2021ybf  & 4.4 & 20.1 & 1.74 & 18.5 & 0.1 & 0.02 & {1.79} &  \\ \hline
\end{tabular}
\caption{Some properties of the 40 LSNe that \cite{Gomezetal2022} study. The first column is the name. Columns 2-5 are properties that \cite{Gomezetal2022} fit to these LSNe by the magnetar and nickel model: the kinetic energy of the ejecta, the ejecta mass, the nickel mass and the initial spin period of the NS. The sixth column lists the magnetar initial energy according to equation (\ref{eq:Emagnetar}) and for a NS mass of $M_{\rm NS}=1.7 M_\odot$. The next two columns list $\eta_{\rm min}$ and $\eta_{\rm max}$, respectively. In the last column I list the problems I find in the magnetar and nickel model that \cite{Gomezetal2022} fit to some of the LSNe: `Excess-E' means that the model predicts energy that is larger than the kinetic energy of the ejecta according to the fitting procedure. `Missing-E' means that the model gives energy that is lower than the kinetic energy.  }
\label{Table:fit}
\end{table*}

For the initial energy of the magnetar \cite{Gomezetal2022} take the value as in MOSFiT
\begin{equation}
\begin{split}
 E_{\rm mag} & = \frac{1}{2} I_{\rm NS} 
 \left( \frac{2 \pi } {P_{\rm spin}} \right)^2 
= 2.6 \times 10^{52} 
\\ & \times 
\left( \frac{M_{\rm NS}} {1.4 M_\odot} \right)^{3/2}
 \left( \frac{P_{\rm spin}}{1 \msec} \right)^{-2} \erg,
\label{eq:Emagnetar}
\end{split}
\end{equation}
where $I_{\rm NS}$ is the moment of inertia of the neutron star of mass $M_{\rm NS}$ and $P_{\rm spin}$ is the initial spin period of the NS. 
In the sixth column of Table \ref{Table:fit} I list the magnetar energy for $M_{\rm NS}=1.7 M_\odot$. 

\section{On the problems of the magnetar and nickel powering} 
\label{sec:Problems}

In this section I point to problems in the energy balance that I find in some of the magnetar and nickel fittings that \cite{Gomezetal2022} find for LSNe. 

\subsection{Fittings with a missing energy } 
\label{subsec:MissingEnergy}

\cite{Gomezetal2022} mention that helium burning to nickel can supply missing energy when a magnetar only model does not supply sufficient energy.  
For a pure helium burning all the way to $^{56}{\rm Ni}$ the energy production is  
\begin{equation}
E_{\rm He,Ni} = 3.3 \times 10^{51} 
\left( \frac{M_{\rm Ni}}{1 M_\odot} \right) \erg. 
\label{eq:Enickel}
\end{equation}
To obtain the maximum kinetic energy that the modelling by \cite{Gomezetal2022} can give I take the magnetar energy (equation \ref{eq:Emagnetar}) for a NS mass of $M_{\rm NS}=1.9 M_\odot$ and add the maximum kinetic energy that the delayed neutrino mechanism can supply of $E_{\rm d-\nu, max} \simeq 2 \times 10^{51} \erg$ (section \ref{sec:intro}). I then find the ratio of this energy to the kinetic energy that \cite{Gomezetal2022} give for their models
\begin{equation}
\eta_{\rm max} \equiv 
\frac{E_{\rm d-\nu, max} + E_{\rm He,Ni} +
E_{\rm mag}(1.9 M_\odot)}{E_{\rm kin}} .
\label{eq:EtaMax}
\end{equation}
I list the values of $\eta_{\rm max}$ in the eighth column of Table \ref{Table:fit}. 

I find that there are eight LSNe out of 40 for which the model cannot supply the kinetic energy that the fitting procedure requires, i.e., I find that $\eta_{\rm max} <1 $. I mark these missing energy cases with `Missing-E' in the last column.  

\subsection{Fitting with extra energy } 
\label{subsec:ExtraEnergy}

If we neglect the energy that the explosion itself supplies the available energy in the fitting procedure of \cite{Gomezetal2022} is the combined energy of nickel formation and the magnetar energy. The radiated energy is small compared to the kinetic energy. If I further neglect the energy source of nickel formation the minimum energy is the magnetar energy for a NS of mass $M_{\rm NS}=1.5 M_\odot$. I define therefore the ratio     
\begin{equation}
\eta_{\rm min} \equiv 
\frac{E_{\rm mag}(1.5 M_\odot)}{E_{\rm kin}} .
\label{eq:EtaMin}
\end{equation}
 I list the values of $\eta_{\rm min}$ in the seventh column of Table \ref{Table:fit}.  
 
I find that there are 10 LSNe with extra energy, i.e., where according to the fitting by \cite{Gomezetal2022} the energy that the magnetar alone supplies is larger than the kinetic energy according to the fitting procedure. I mark these with `Excess-E'. 

It seems that a value of $\eta_{\rm min}>1$ points to inconsistency in the fitting procedure with a magnetar as the main extra (extra to the explosion itself) energy source.
{ In that regards I note that it is not possible according to the magnetar model that a small ejecta mass that avoids detection takes the excess energy that the magnetar supplies in these ten LSNe. The reason is that the magnetar modelling of the extra energy assumes that the energy that the magnetar releases is thermalized in the ejecta. A fraction of this energy is radiated away. Any mass that expands at a very high velocity would take energy from this thermalization and as a consequence from radiation. Therefore the model would not be able to account for the light curve. 
}

\subsection{Comments} 
\label{subsec:Oxugen}

First I comment on the error bars of the fitted parameters. There are large error bars on the parameters that \cite{Gomezetal2022} fit. The largest relevant error bars are in the kinetic energies in their derivation of parameters. Crudely, a typically error bar on the kinetic energy allows up to twice as large kinetic energy as the given value and down to half the given value. Even if I include all error bars, there are still six systems with excess energy (instead of ten).

Referring to these error bars does not change the conclusion here. I here examine whether the best fitted model is consistent with energy considerations. As one can see from the values of $\eta_{\rm max}$ of the eight LSNe that I mark to have missing energy (Missing-E in the last column of Table \ref{Table:fit}), five LSNe have $\eta_{\rm max} < 0.7$. And I recall that in calculating $\eta_{\rm max}$ I took the maximum possible energy of the neutrino driven mechanism. Therefore, although the large error bars include the ratio of $\eta_{\rm max}=1$, I claim that the modelling is not fully consistent.  
Out of 10 cases with $\eta _{\rm min} >1$, six have excess energy even when the error bars are included. But here again, the error bars are less relevant in checking the consistency of the model. 
 
A consistent fitting should not allow values of neither $\eta_{\rm max}$ below unity nor $\eta_{\rm min}$ above unity. My argument is that powering by jets must be allowed in the fitting procedure (section \ref{sec:Jets}).    

My second comment refers to the claim of helium burning to nickel-56. 
\cite{Gomezetal2022} notice that the magnetar powering alone cannot explain many of the LSNe that they study. For that, and because they avoided using jets, they mention the extra energy of pure helium burning to $^{56}{\rm Ni}$. In massive cores there is a massive oxygen shell between the silicon layer and the helium layer. Therefore, it is not clear how the explosion causes the helium to burn to nickel, but does not ignite several solar masses of oxygen to burn to nickel. I take this problem to be severe when according to the fitting by \cite{Gomezetal2022} the nickle mass is $M_{\rm Ni}>0.5 M_\odot$ and the ejected mass is $M_{\rm ej} > 5 M_\odot$. There are 12 such LSNe. However, because it is not clear in which of these LSNe the entire nickel is nucleosynthesised from helium, I will not discuss this problem further. I only stress my claim that the extra energy comes mainly from jets and not from helium burning to $^{56}{\rm Ni}$. 

Overall, there are 18 out of 40 LSNe where I find problems with the MOSFiT fitting by \cite{Gomezetal2022}. This finding strengthens my earlier claim that SLSNe, and now LSNe as well, are powered by jets and that any model with energetic magnetar cannot ignore jets 
\citep{Soker2016Magnetar, Soker2017Magnetar, SokerGilkis2017}. I therefore turn to further explore the nature of jets in SLSNe and LSNe. 

\section{Powering by jets} 
\label{sec:Jets}

\subsection{A toy model} 
\label{subsec:ToyModel}

The influence of jets on the light curve is impossible to model by a simple formula because of the huge volume of the parameter space, e.g., the duration of the jet-active phase, the mass loss rate into the jets and its time variation, the opening angle of the jets, the explosion morphology, the ejecta mass, the explosion energy, and the viewing angle. Most problematic to model is the power of the post-explosion jets as function of time $P_{\rm jet}(t)$. Each case needs its own fitting.

\cite{KaplanSoker2020b} modelled the light curve of SN~2018don, one of the 40 LSNe that \cite{Gomezetal2022} study. \cite{KaplanSoker2020b} described the case where strong jets power a bipolar explosion and argued that for an observer in the equatorial plane of the bipolar ejecta the light curve has a rapid luminosity decline. Then there is a `knee' where the decline becomes more moderate. They then used this model to fit the light curve of SN~2018don that has an abrupt decline in its light curve. Note that they did not include very late (post-explosion) jets in their modelling, but rather jets at the explosion (or very shortly thereafter, much less than the rise time).   
I here take a different approach. Building on the toy model of \cite{KaplanSoker2020a} of late jets I crudely estimate the relation between the maximum rise time (for a given luminosity) and the luminosity. 

\cite{KaplanSoker2020a} built a toy model to explore the influence of late jets on the light curve of CCSNe. They assumed that the jet activity phase lasts for a very short time, and considered the short interaction of each of the two jets with the CCSN ejecta to be an off-center `mini-explosion'. There is one mini-explosion on each side of the equatorial plane, as there are two opposite jets. 
This interaction shocks a region of the ejecta to form the `cocoon', one cocoon for each of the two opposite jets. Using simple analytical formulae \cite{KaplanSoker2020a} estimated the peak extra luminosity $L_{\rm c}$ and the peak typical time scale $t_{\rm c}$. 
Their equations read 
\small
\begin{eqnarray}
\begin{aligned} 
& L_{\rm c} = 4.2\times 10^{41} 
\left(\frac{\epsilon_V}{0.067}\right)^{-1}
\left(\frac{\epsilon_E}{0.01}\right) \left(\frac{\sin{\alpha_{\rm j}}}{0.087}\right) \\ & \times
\left(\frac{\beta}{0.5}\right)
\left(\frac{M_{\rm SN}}{10 M_{\odot}}\right)^{-3/2}
\left(\frac{E_{\rm SN}}{2\times 10^{51} \erg}\right)^{3/2}
 \\ & \times
\left(\frac{\kappa_{\rm c}}{0.38  \cm^2\g^{-1}}\right)^{-1}
\left(\frac{t_{\rm j,0}}{100 {~\rm d}}\right) \erg \s^{-1}, 
\end{aligned}
\label{eq:Ljet}
\end{eqnarray}
\normalsize
and
\small
\begin{eqnarray}
\begin{aligned} 
& t_{\rm c} = 56 
\left(\frac{\epsilon_V}{0.067}\right)^{3/4}
\left(\frac{\epsilon_E}{0.01}\right)^{-1/4}
\left(\frac{E_{\rm SN}}{2\times 10^{51} \erg}\right)^{-1/4} \\ & \times
\left(\frac{M_{\rm SN}}{10 M_{\odot}}\right)^{3/4}
\left(\frac{\kappa_{\rm c}}{0.38 \cm^2\g^{-1}}\right)^{1/2} {~\rm d}, 
\end{aligned}
\label{eq:tjet}
\end{eqnarray}
\normalsize
respectively. In these equations 
$\epsilon_V$ is the mass ratio in one cocoon to the total CCSN ejecta mass, 
$\epsilon_E$ is the ratio of the energy that one jet deposits into one cocoon to the total CCSN explosion energy,  
$\alpha_{\rm j}$ is the half opening angle of the jet, 
$\beta$ is the ratio of the radius at which the `mini-explosion' occurs to the ejecta outer radius, $M_{\rm SN}$ is the ejecta mass in the explosion, $E_{\rm SN}$ is the kinetic energy of the ejecta, $\kappa_{\rm c}$ is the opacity, and $t_{\rm j,0}$ is the time of the `mini-explosion' after the CCSN explosion. 

To reach my goal of obtaining the maximum rise time due to energetic late jets I make the following minimum necessary changes from their toy model. These of course include a very simple and crude assumptions.  

(1) I assume that the longest rise time are determined by a long jet-activity phase. These jets are powered by fall back material that feeds an accretion disk around the newly born NS (or black hole even). The interaction time starts early and ends at a final fime of $t_{\rm j,f}$. 

(2) I keep the explosion energy and ejecta mass as in the scaling of \cite{KaplanSoker2020a} (equations \ref{eq:Ljet} and \ref{eq:tjet}). For long rise times in this modelling the ejecta mass must be large. I keep it at $M_{\rm SN}=10 M_\odot$, but note that it might be larger even. I also keep the explosion energy itself at $E_{\rm SN}=2 \times 10^{52} \erg$, but note that in the jet-driven explosion mechanism the energy might be larger even. A massive ejecta which implies a massive core in CCSNe-I (stripped-envelope SNe) might account for the long accretion phase of fallback material.  

(3) I change the opacity to $\kappa_{\rm c} =0.1 \cm^2 \g^{-1}$, as a typical value that \cite{Gomezetal2022} list, although they also fit LSNe with opacity values down to $0.02 \cm^2 \g^{-1}$. 

(4) I assume that the jets are active over a period of time, and so I take the energy of the jets to increase linearly with time. I scale the total jet's energy with 
\begin{equation}
{\epsilon_E}={\epsilon_{E,0}} \left( \frac{t_{\rm j,f}}{100 {~\rm d}} \right), 
\label{eq:EpsilonE}
\end{equation}
and with $\epsilon_{E,0}=1$, where ${\epsilon_E} E_{\rm SN}$ is the total energy of one jet out of the two opposite jets, and the jets can be active for more than 100 days. 

(5) I assume that the jets-ejecta interaction moves outward with time, and take (for simplicity) a linear relation for the radius of interaction relative to the radius of the ejecta 
\begin{equation}
\beta = \beta_0 \left( \frac{t_{\rm j,f}}{100 {~\rm d}} \right).
\label{eq:Beta0}
\end{equation}
This implies that this relation holds to $t_{\rm j,f} < 100 {~\rm d}/\beta_0$. In any case, I will take $\beta_0=0.5$ and will not consider times much beyond $t_{\rm j,f}=130~{\rm d}$.    

I will take the jets to be very energetic, so that they dominate the emission. Namely, the luminosity of the light curve is the luminosity due to the late jets interaction with the ejecta. I also ignore the effects of the viewing angle despite that the ejecta with such strong jets must be bipolar. 
With these assumptions equation (\ref{eq:Ljet}) becomes  
\small
\begin{eqnarray}
\begin{aligned} 
& L_{\rm c} \simeq 1.6\times 10^{44} 
\left(\frac{\epsilon_V}{0.067}\right)^{-1}
\left(\frac{\epsilon_{E,0}}{1}\right)
\left(\frac{\sin{\alpha_{\rm j}}}{0.087}\right) \\ & \times
\left( \frac{\beta_0}{0.5} \right)
\left(\frac{M_{\rm SN}}{10 M_{\odot}}\right)^{-3/2}
\left(\frac{E_{\rm SN}}{2\times 10^{51} \erg}\right)^{3/2}
 \\ & \times
\left(\frac{\kappa_{\rm c}}{0.1  \cm^2\g^{-1}}\right)^{-1}
\left(\frac{t_{\rm j,f}}{100 {~\rm d}}\right)^3 \erg \s^{-1}. 
\end{aligned}
\label{eq:Ljetmax}
\end{eqnarray}
\normalsize

\begin{figure}
 \centering
\includegraphics[trim= 2.6cm 14.8cm 0.0cm 1.5cm,scale=0.53]{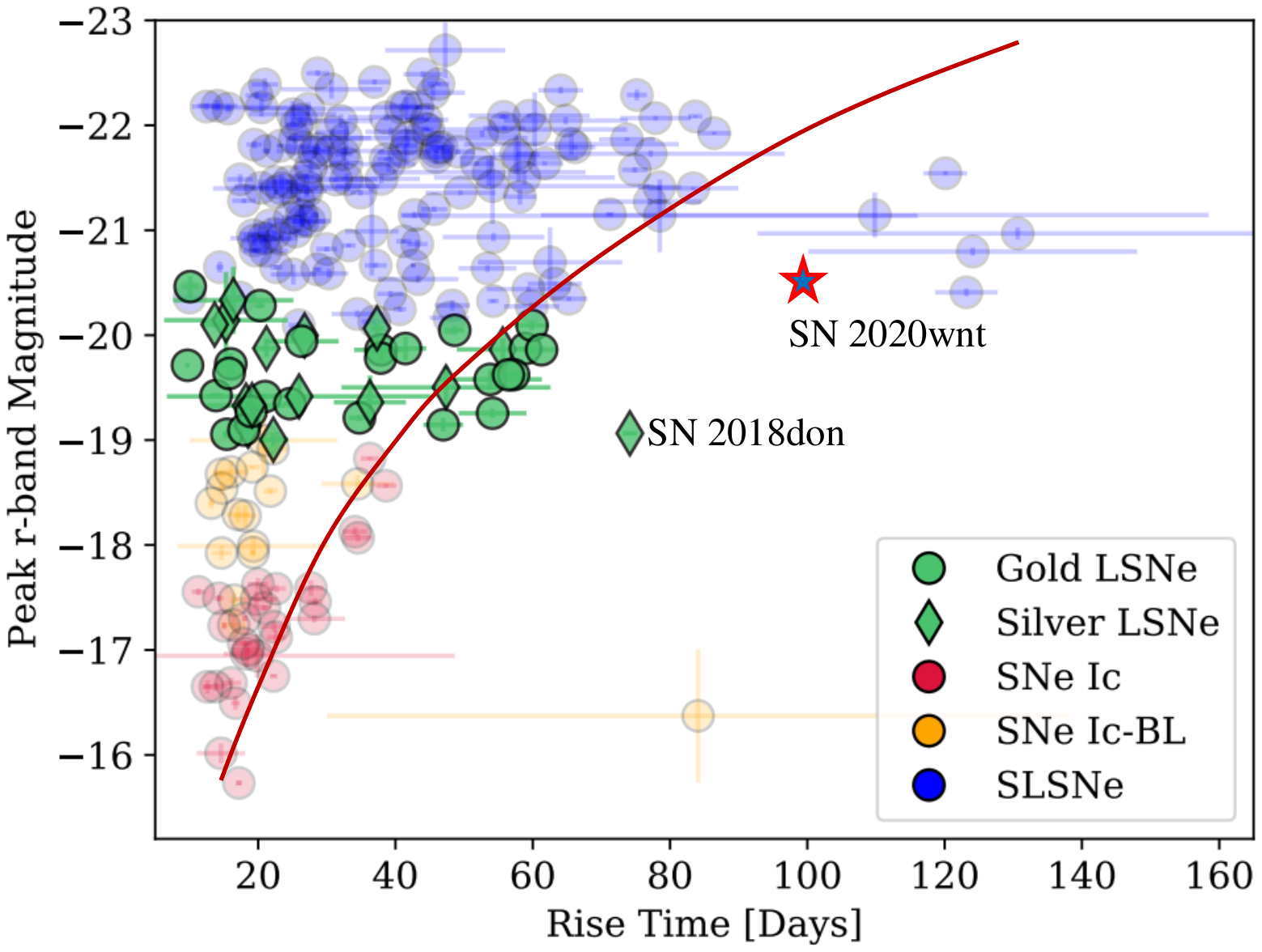}
\caption{A figure from \cite{Gomezetal2022} of CCSNe-I in the plane of peak r-band absolute magnitude versus luminosity rise time. The inset lists the different CCSN classes. I added SN~2020wnt and the red line according to equation (\ref{eq:Ljetmax}). }
 \label{Fig:MrRiseTime}
 \end{figure}

The rise time might be somewhat longer than $t_{\rm j,f}$ due to photon diffusion time, but here I take the rise time to be the end time of the activity phase of the powerful jets, i.e., $t_{\rm rise} \simeq t_{\rm j,f}$. 
For the absolute r-band magnitude I take 
$M_{\rm r}= 4.64 - 2.5 \log (L_{\rm c}/L_\odot)$. 
In Fig. \ref{Fig:MrRiseTime} I draw the r-band magnitude as I calculate from equation (\ref{eq:Ljetmax}) as function of the rise time that I take to be $t_{\rm j,f}$ with a solid-red line on top of a figure from \cite{Gomezetal2022}. 

\subsection{Interpretation} 
\label{subsec:Interpretation}

In section \ref{subsec:ToyModel} I extended the late-jets toy model of \cite{KaplanSoker2020a} to energetic jets with a long activity time period. I draw the relation between the peak r-band magnitude and the rise time according to the scaling of equation(\ref{eq:Ljetmax}) by a red line on a figure from \cite{Gomezetal2022}. I now interpret this figure.

Firstly, I note that one LSN and four SLSNe in the original plot of \cite{Gomezetal2022} have much longer rise times that the others.
I also added the stripped-envelope SN~2020wnt (\citealt{Tinyanontetal2021, Gutierrezetal2022, Tinyanontetal2022}). These require a different modelling, and probably have a very massive ejecta. I discuss them in section \ref{subsec:SN2018don}. 

Because the very simple toy model and the very large parameter space for the different variables and their variations with time and their variations from one LSN to another, the red line on Fig. \ref{Fig:MrRiseTime} represents only a trend, and not a fit. The trend it represents is that for massive ejecta, $M_{\rm SN} \ga 10 M_\odot$, which through a fall back accretion allow a long jet-activity phase, the peak luminosity (as given by equation \ref{eq:Ljetmax}) increases with a large power of the jet-activity time $L_c \propto t^{\xi}_{\rm j,f}$ with $\xi \simeq 3$.  
For a given jet-activity duration the peak luminosity can be much higher if the ejecta mass is smaller and/or, for example, the jets themselves are more energetic, i.e., $\epsilon_{E,0}$ is larger.  Again, I cannot justify the specific scaling of all variable. I took some as the scaling that \cite{KaplanSoker2020a} use (as I described above), and for others I substituted plausible average values ($\kappa_{\rm c}$ and equations \ref{eq:EpsilonE} and \ref{eq:Beta0}). 

For a given peak luminosity, and keeping other parameters in equation (\ref{eq:Ljetmax}) unchanged, the rise time is maximum for maximum ejecta mass $t_{\rm rise} \simeq t_{\rm j,f} \propto M^{1/2}_{\rm SN}$. However, as mentioned above the other parameters can vary a lot from one object to the next.  In addition, the viewing angle might play a significant role. 

For CCSNe to the left of the red line the jet activity phase is shorter (or the power of the jets rapidly decreases with time). For these CCSNe it is possible that the powering is only by jets launched at the explosion itself. And again, I do not claim that a magnetar does not supply extra energy as well. I rather claim that jets must be included when modelling with an energetic magnetar.  

The main point of this section is that the late-jet modelling might account for the general boundary of most CCSNe-I (hydrogen deficient CCSNe) in the plane of the peak r-band magnitude versus rise time.

\subsection{SN~2018don} 
\label{subsec:SN2018don}

There is one LSN on Fig. \ref{Fig:MrRiseTime} that is off to the right with respect to other LSNe. This is SN~2018don. As well, it does not fit to the present extension (section \ref{subsec:ToyModel}) of the toy model of \cite{KaplanSoker2020a}. 

\cite{KaplanSoker2020b} modelled the light curve of SN~2018don with jet-driven bipolar explosion, rather than with long-lived jets (any delay of the jets after explosion is much shorter than the rise time). The bipolar explosion model is probably not unique. However, it does fit the rapid (almost abrupt) drop in the light curve of SN~2018don (for the light curve see \citealt{Lunnanetal2020}). This modelling requires both an equatorial observer and a highly bipolar explosion, which together account for such a behavior being rare.  

\cite{Gomezetal2022} note that \cite{Lunnanetal2020} consider SN~2018don to be a SLSN (rather than a LSN). \cite{Gomezetal2022} comment that if indeed its host galaxy has a substantial extinction then its peak luminosity is higher and it is a SLSN-I. In that case it will not be much off to the right relative to the other CCSNe-I in Fig. \ref{Fig:MrRiseTime}.  

SN~2020wnt  has a peak magnitude of $-20.5$ \citep{Tinyanontetal2021, Gutierrezetal2022} and a rise time of $\simeq 80$~days \citep{Gutierrezetal2022} to possibly $\simeq 100$~days \citep{Tinyanontetal2022}, depending on the time of explosion. I added SN~2020wnt to Fig. \ref{Fig:MrRiseTime}. I suggest that a jet-driven energetic bipolar explosion accounts also for the behavior of SN~2020wnt, including the knee in its light curve \citep{Tinyanontetal2022}, something the bipolar explosion model can account for \citep{KaplanSoker2020b}.  

The point I emphasize here is that jets introduce a very large parameter space that includes also rare combinations of parameters. Some of these rare combinations might account for SN~2018don, SN~2020wnt, and the other five SLSNe that are off to the right in Fig. \ref{Fig:MrRiseTime}.  

\section{Summary} 
\label{sec:Summary}

Adopting the view that jets power most, or even all, CCSNe, i.e., the jittering jets explosion mechanism (section \ref{sec:intro}), and motivated by the finding that SLSNe cannot be powered by the delayed neutrino mechanism plus magnetar alone \citep{SokerGilkis2017}, I set to examine the modelling by \cite{Gomezetal2022} of 40 LSNe with extra powering by a magnetar and nickel, but without the inclusion of jet-powering (Section \ref{sec:Magnetar}). 

I find that 18 fittings out of the 40 fittings to these LSNe suffer from a problem. I list these problems in the last column of Table \ref{Table:fit}. 
 
I find (Section \ref{subsec:MissingEnergy}) that the fitting procedure by \cite{Gomezetal2022} to eight LSNe yields an ejecta  kinetic energy that is larger than what the combined energy of the delayed neutrino mechanism, a magnetar, and helium burning to nickel can supply. I mark these with `Missing-E' in the last column of Table \ref{Table:fit}. Clearly, an extra energy source is needed, which I claim to be jets. 

I find (section \ref{subsec:ExtraEnergy}) that for 10 LSNe the fitting procedure by \cite{Gomezetal2022} gives a magnetar energy that is larger than the ejecta kinetic energy according to the fitting. I mark these with `Excess-E' in the last column of Table \ref{Table:fit}. My explanation is that jets supply a substantial amount of energy rather than these two sources. Such energetic jets might lead to a bipolar explosion, and therefore the viewing angle is also an important parameter in the fitting. \cite{KaplanSoker2020b} demonstrated the possible importance of the viewing angle in the their bipolar explosion model of SN~2018don (section \ref{subsec:SN2018don}). I argued in section \ref{subsec:SN2018don} that a similar model of energetic jet-driven bipolar explosion (that works even without late jets) account also for SN~2020wnt. 

In total I find that this fitting procedure that ignores jets encounters problems in 18 out of the 40 LSNe. I therefore strengthen my claim \citep{Soker2016Magnetar, Soker2017Magnetar, SokerGilkis2017} that jets launched at magnetar birth cannot be ignored. 

In section \ref{sec:Jets} I extended the toy model of \cite{KaplanSoker2020a} to include late post-explosion jets that are active for weeks to months. With this very simple toy model I draw the red line on Fig. \ref{Fig:MrRiseTime}. This line is more or less along the boundary of most CCSNe-I in the plane of peak r-band absolute magnitude versus rise time. This line is not a fit nor a complete model. It only serves to indicate that long-lasting post-explosion jets might account for some properties of CCSNe, in particular LSNe and SLSNe. Those CCSNe to the far right of the red line require different modelling, as I discussed in section \ref{subsec:SN2018don}. 

The newly born NS can launch late jets as it accretes fallback ejecta material (see, e.g., \citealt{Pellegrinoetal2022} for a recent discussion of some SNe Icn). Alternatively, it can accrete from a main sequence companion as \cite{Hoberetal2022} study. 
The mass removal from the progenitors of stripped-envelope CCSNe is likely to be due the presence of a close companion that performs a common envelope evolution. If the companion is a main sequence star that ends at $\simeq 10-20 R_\odot$ at explosion then the collision of the CCSN ejecta with the main sequence star inflates its envelope \citep{Ogataetal2021}. The main sequence might engulf then the newly born NS. The NS spirals-in inside the inflated envelope and launches jets as it accretes mass via an accretion disk. \cite{Hoberetal2022} argue that the jets might last for a time period from few weeks to several months.  

I end by reiterating my claims that jets launched at magnetar birth cannot be ignored, and that jets supply most of the energy of LSNe and SLSNe, and possibly most of the energy in the majority of CCSNe. 

\section*{Acknowledgments}

I thank Avishai Gilkis, Dmitry Shishkin, and an anonymous referee  for helpful comments. 
This research was supported by a grant from the Israel Science Foundation (769/20).



\end{document}